\documentclass[aps,preprint,showpacs]{revtex4}
\usepackage{graphicx}

\begin{document}

\title{Broadband wide-angle absorption enhancement due to mode conversion in cold unmagnetized plasmas with periodic density variations}

\author{Dae Jung \surname{Yu}}
\affiliation{Center for Mathematical Plasma Astrophysics, Department of Mathematics, KU Leuven,
Celestijnenlaan 200B bus 2400, B-3001 Leuven, Belgium}
\author{Kihong \surname{Kim}}
\email{khkim@ajou.ac.kr}
\affiliation{Department of Energy Systems
Research and Department of Physics, Ajou University, Suwon 16499, Korea}

\begin{abstract}
We study theoretically the mode conversion and the associated resonant absorption of $p$-polarized electromagnetic waves into longitudinal plasma oscillations in cold, unmagnetized and stratified plasmas with periodic spatial density variations. We consider sinusoidal density configurations for which the frequency band where mode conversion occurs is well included within a transmission band of the one-dimensional plasma photonic crystal. We calculate the mode conversion coefficient, which measures the fraction of the electromagnetic wave energy absorbed into the plasma, and the spatial distribution of the magnetic field intensity for various values of the wave frequency and the incident angle using the invariant imbedding theory of mode conversion. We find that the absorption is greatly enhanced over a wide range of frequency and incident angle due to the interplay between the mode conversion and the photonic band structure. The enhancement occurs because for frequencies within a transmission band, the wave reflection is strongly suppressed and the waves penetrate more deeply into the inhomogeneous region, thereby increasing the possibility for them to reach many resonance points where the dielectric permittivity vanishes.
\end{abstract}


\maketitle

\section{Introduction}

Photonic crystals are the structures in which some electromagnetic constitutive parameters vary periodically in space \cite{1,2,3}.
These structures exhibit photonic band gaps in their frequency spectra due to the interference of multiply-scattered waves, which influence the wave propagation characteristics enormously. Photonic crystals have played a central role in the modern development
of optics and photonics.

More recently, the photonic crystal concept has been extended to plasma physics. Plasma photonic crystals (PPCs) refer to plasma media with periodic spatial density variations or periodic composite structures of plasmas and solids.
Periodic modulations of the plasma density can be induced in plasma-filled backward-wave oscillators by the ponderomotive potential effect,
which is due to the strong interaction between plasmas and two counter-propagating electromagnetic waves \cite{4,5}.
It is also possible to create PPCs using the dielectric barrier discharge method \cite{6,7}.
The wave propagation characteristics of various types of one-dimensional and two-dimensional PPCs and their applications have been studied extensively by a number of researchers \cite{8,9,10,11,12,13}.

One interesting phenomenon concerning the propagation of electromagnetic waves in inhomogeneous plasmas is the mode conversion of one type of wave mode into another type in some regions where the two modes meet the resonance condition \cite{14,15,16}. The irreversible transfer of electromagnetic wave energy to the resonance region associated with the mode
conversion occurring in unmagnetized or magnetized inhomogeneous plasmas plays a central role in a wide
range of phenomena in plasma physics \cite{17,18,19,20,21}.
We are mainly interested in the mode conversion of transverse electromagnetic waves into longitudinal plasma oscillations in cold unmagnetized plasmas, which occurs at the regions where the local dielectric permittivity vanishes, or equivalently, the wave frequency matches the local plasma frequency \cite{18,19,22,23,24}. In this paper, we consider plasma density configurations where this resonance condition is satisfied periodically along one direction in space. The interplay of photonic band gap and mode conversion in such systems has never been investigated before. We explore this phenomenon by calculating
the mode conversion coefficient, which measures the wave absorption,
in a numerically exact manner using the invariant imbedding method \cite{19,20,25,26,27}.

In a given configuration of the plasma density, the mode conversion is allowed to occur within a certain frequency band, which we call the mode conversion band.
We find that when the mode conversion band is well within the transmission band of a PPC, the mode conversion coefficient is greatly enhanced for a broad frequency range and for a wide range of incident angle. This occurs because the wave reflection is strongly suppressed for frequencies in the transmission band and the wave is able to penetrate more deeply into the periodic region, thus enhancing the efficiency of mode conversion.

In Sec.~\ref{sec2}, we introduce our model with sinusoidal density variations. In Sec.~\ref{sec3}, we give a brief summary of the invariant imbedding method used in this paper. In Sec.~\ref{sec4}, we present our numerical results on the dependencies of the mode conversion coefficient on the frequency and the incident angle as well as the spatial field distribution in some special cases. Finally, in Sec.~\ref{sec5}, we summarize the paper.

\section{Model}
\label{sec2}

Since in cold unmagnetized plasmas the mode conversion occurs only for $p$-polarized waves, we consider only that case.
We are interested in the propagation of a $p$ wave of frequency
$\omega$ and vacuum wave number $k_0$ ($=\omega/c$) in a stratified, cold, and unmagnetized plasma, where the electron density $n$ and the
dielectric permittivity $\epsilon$ vary only along the $z$ axis. The
inhomogeneous plasma lies in $0 \le z \le L$ and the wave propagates
in the $xz$ plane.
In a cold unmagnetized plasma, $\epsilon$ is written as
\begin{equation}
\epsilon(z)=1-\frac{\omega_p^2}{\omega(\omega+i\gamma)},~~ \omega_p^2=\frac{4\pi e^2}{m}n(z),
\end{equation}
where $\gamma$ is the collision frequency and $\omega_p$ is the electron
plasma frequency. $e$ and $m$ are the electron charge and
mass respectively. We assume that in the region $0\le z\le L$, the electron density $n(z)$ varies sinusoidally as
\begin{equation}
 n(z) = n_0\left[1+u\sin{\left(2\pi \frac{L-z}{\Lambda}\right)}\right],
\end{equation}
where $u$ ($0<u\le 1$) and $\Lambda$ are the strength and the spatial period of the sinusoidal variation.
Outside the inhomogeneous region, the electron density is equal to $n_0$ and the collision frequency is assumed to be zero.
The corresponding expression for $\epsilon$ is
\begin{widetext}
\begin{equation}
 \epsilon(z) = \left\{ \begin{array}{ll}
     1-\frac{\Omega^2}{\omega^2} & \mbox{ if $z> L$, $z<0$}\\
       1-\frac{\Omega^2}{\omega^2}\left[1+u\sin{\left(2\pi \frac{L-z}{\Lambda}\right)}\right]+ i\eta & \mbox{ if $0\leq z\leq L$ }\end{array}  \right.,
\end{equation}
\end{widetext}
where $\Omega^2=4\pi n_0e^2/m$ and $\eta$ is the dimensionless damping parameter proportional to
$\gamma$, which we will choose to be extremely small. In order to have a propagating wave outside the inhomogeneous region, $\omega/\Omega$ has to be greater than 1.

In Fig.~1, we illustrate the typical spatial dependence of the electron plasma density by plotting $n(z)/n_0$ versus
$z/\Lambda$ when $u=0.6$ and $L/\Lambda=20$. The maximum value of the plasma frequency, $\omega_p^{\rm max}$, is given by
$\sqrt{1+u}~\Omega$, which is approximately equal to $1.265\Omega$ when $u=0.6$.

\begin{figure}
\centering
\includegraphics[width=3.2in]{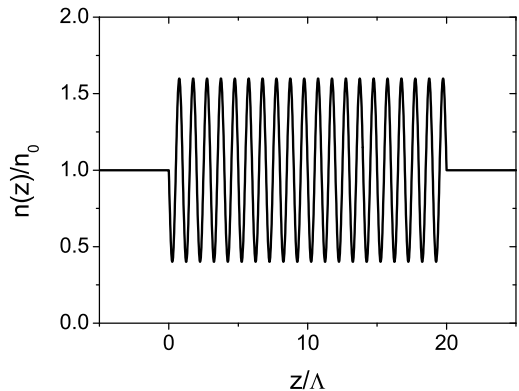}
\caption{Spatial dependence of the normalized electron density, $n(z)/n_0$, when $u=0.6$ and $L/\Lambda=20$. $\Lambda$
is the spatial period of the sinusoidal variation.}
\end{figure}

\section{Invariant imbedding method}
\label{sec3}

In this section, we give a brief summary of the invariant imbedding method used in the present study.
More details can be found in \cite{19,26}.
For obliquely incident $p$ waves, the complex amplitude of the magnetic field satisfies
\begin{equation}
\frac{d^2 H}{d z^2}
   - \frac{1}{\epsilon(z)}\frac{d \epsilon}{d z}\frac{d H}{d z}
      + [k_{0}^2 \epsilon(z)-q^2]H = 0,
      \label{eq:pe1}
\end{equation}
where $q$ is the $x$ component of the wave vector. When $\theta$ is
the angle of incidence, $q$ is equal to $\sqrt{\epsilon_i}k_0\sin\theta$, where $\epsilon_i$ ($=1-\Omega^2/\omega^2$) is the dielectric permittivity in the incident region.

We consider a $p$ wave of unit magnitude incident
on the plasma from the region where $z>L$ and transmitted to the region where $z<0$. The quantities of main interest are the
complex reflection and transmission coefficients, $r=r(L)$ and $t=t(L)$, which we consider as functions
of the thickness $L$.
Using the invariant imbedding method, we derive exact
differential equations satisfied by $r$ and $t$:
\begin{eqnarray}
 &&\frac{1}{p}\frac{dr}{dl}=2i\epsilon{(l)}r
-\frac{i}{2}
\Bigl[\epsilon(l)-1\Bigl]\Bigl[1-\frac{\tan^2{\theta}
}{\epsilon(l)}\Bigl](1+r)^2,\nonumber\\
&& \frac{1}{p}\frac{dt}{dl}=i\epsilon{(l)}t
-\frac{i}{2}
\Bigl[\epsilon(l)-1\Bigl]\Bigl[1-\frac{\tan^2{\theta}
}{\epsilon(l)}\Bigl](1+r)t,
\label{eq:imbeq}
\end{eqnarray}
where $p=\sqrt{\epsilon_i}k_0\cos\theta$.
These equations are integrated from
$l=0$ to $l=L$ numerically, using the initial conditions $r(0)=0$ and $t(0)=1$.
When mode conversion happens, there occurs resonant energy absorption of electromagnetic
waves even for a negligibly small damping parameter $\eta$. The
mode conversion coefficient $A$, which measures the wave absorption, is defined by
$A=1-R-T$, where $R$ ($=\vert r\vert^2$) and $T$ ($=\vert t\vert^2$) are the reflectance and the transmittance respectively.

The invariant imbedding method can also be used in calculating the field amplitude $H(z;L)$
for $0\le z\le L$, which we consider as a function of both $z$ and $L$.
The equation satisfied by $H(z;L)$ is very similar to that for $t$ and takes the form
\begin{eqnarray}
&&\frac{1}{p}\frac{\partial}{\partial l}H(z;l)=i\epsilon{(l)}H(z;l)\nonumber\\
&&~~~~-\frac{i}{2}
\Bigl[\epsilon(l)-1\Bigl]\Bigl[1-\frac{\tan^2{\theta}
}{\epsilon(l)}\Bigl]\left[1+r(l)\right]H(z;l),
\label{eq:fe}
\end{eqnarray}
which is integrated from $l=z$ to $l=L$ using the initial condition $H(z;z)=1+r(z)$
to obtain $H(z;L)$.

\begin{figure}
\centering
\includegraphics[width=3.2in]{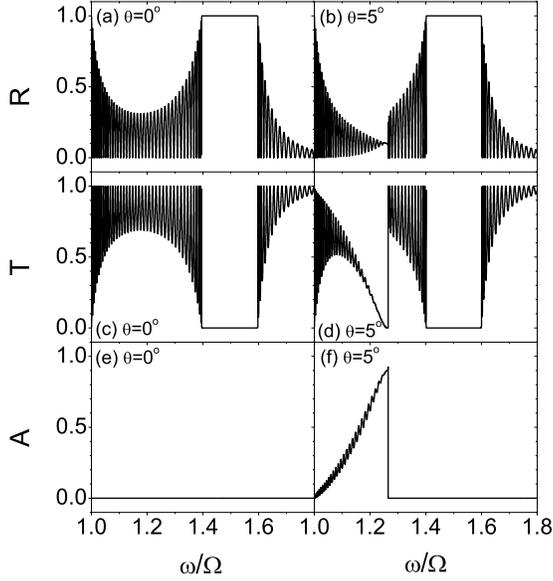}
\caption{Reflectance $R$, transmittance $T$, and mode conversion coefficient $A$ plotted versus the normalized frequency $\omega/\Omega$ for $\theta=0^\circ$ and $5^\circ$, when $u=0.6$, $\eta=10^{-8}$, $L/\Lambda=50$ and $\Omega\Lambda/c=2.8$. }
\end{figure}

\section{Numerical results}
\label{sec4}

\begin{figure}
\centering
\includegraphics[width=3.2in]{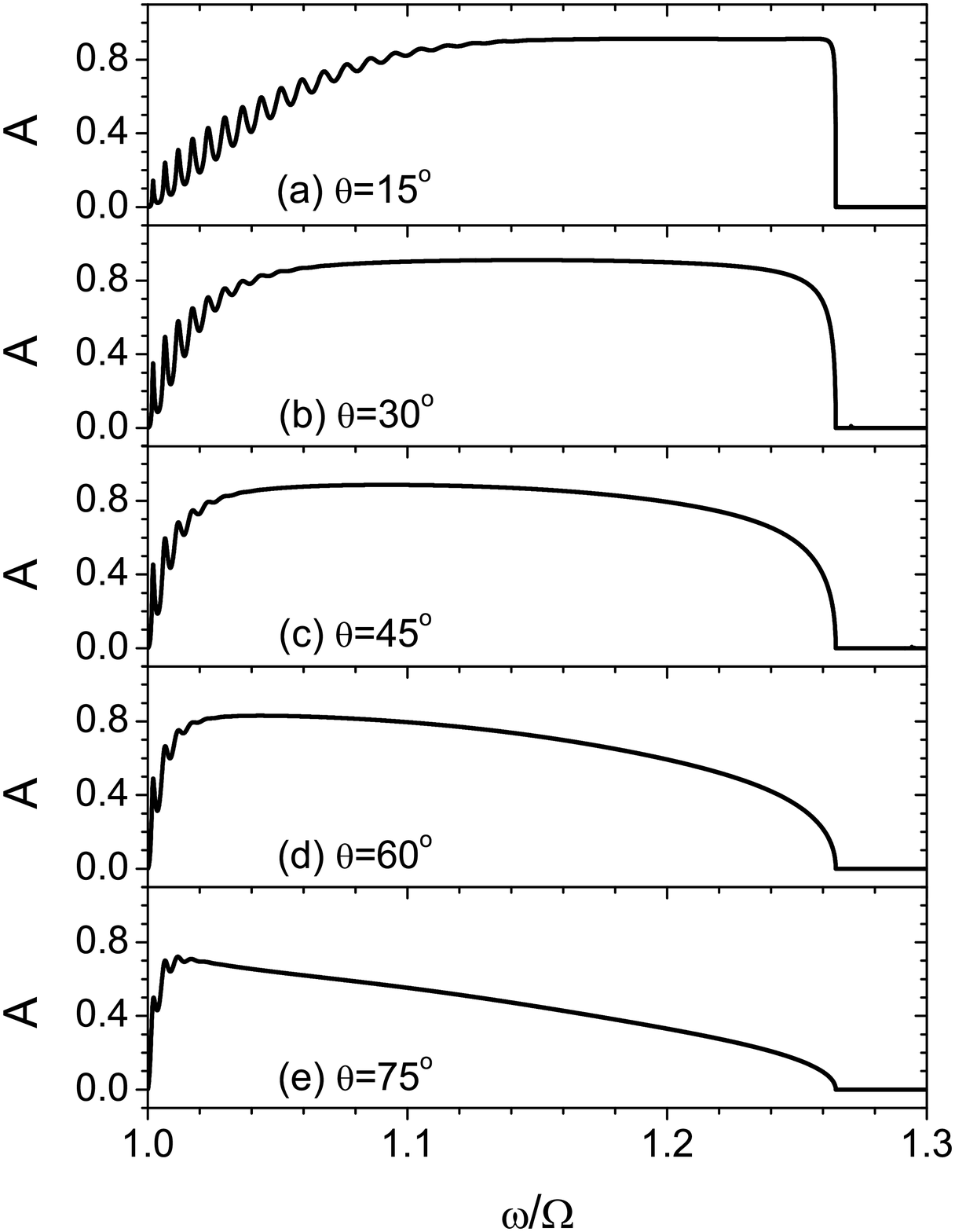}
\caption{Mode conversion coefficient $A$ plotted versus the normalized frequency $\omega/\Omega$ for $\theta=15^\circ$, $30^\circ$, $45^\circ$, $60^\circ$ and
$75^\circ$, when $u=0.6$, $\eta=10^{-8}$, $L/\Lambda=50$ and $\Omega\Lambda/c=2.8$. }
\end{figure}

In Fig.~2, we plot the reflectance $R$, the transmittance $T$ and the mode conversion coefficient $A$ versus the normalized frequency $\omega/\Omega$ for $\theta=0^\circ$ and $5^\circ$, when $u=0.6$, $\eta=10^{-8}$, $L/\Lambda=50$ and $\Omega\Lambda/c=2.8$. The mode conversion is possible if there exists a region where the real part of $\epsilon$ vanishes. In the present configuration, this condition imposes that there is a mode conversion band, or an absorption band, in the frequency region $1<\omega/\Omega< \sqrt{1+u}$. When $u=0.6$, the absorption band is in $1<\omega/\Omega<1.265$, as shown in Fig.~2(f). We have verified numerically that the results do not change as $\eta$ becomes smaller. When mode conversion occurs, the mode conversion coefficient is finite even in the presence of an infinitesimally small amount of damping at the resonance region,
which signifies that the absorption is not due to any dissipative damping process but due to mode conversion. With the choice of $\Omega\Lambda/c=2.8$, we find that there are a transmission band in $1<\omega/\Omega< 1.4$ and a reflection band in $1.4<\omega/\Omega< 1.6$.
In fact, we have chosen the parameter $\Omega\Lambda/c=2.8$ deliberately so that the mode conversion band is well-included in the transmission band.
For normal incidence, mode conversion does not occur, whereas for finite $\theta$, it occurs in the expected frequency region, as can be seen in Fig.~2(f). We note that when this happens, both the reflectance and the transmittance are suppressed.

In Fig.~3, we plot the mode conversion coefficients obtained when the incident angles are $15^\circ$, $30^\circ$, $45^\circ$, $60^\circ$ and
$75^\circ$ versus the normalized frequency $\omega/\Omega$ for the same parameter values as used in Fig.~2. We find that a substantial amount of mode conversion occurs over the full frequency range of the mode conversion band. For instance, when $\theta=30^\circ$, $A$ is larger than 0.5 over almost the entire range of the mode conversion band and larger than 0.8 in a very broad frequency region corresponding to $1.041<\omega/\Omega<1.252$. In the case where the mode conversion in cold unmagnetized plasmas occurs at only one value of $z$ in space, it has been well-known that the maximum value of $A$ is about 0.5 \cite{18,19}. Therefore, the present result confirms that mode conversion is greatly enhanced due to the fact that it occurs within a transmission band of the PPC.

\begin{figure}
\centering
\includegraphics[width=3.2in]{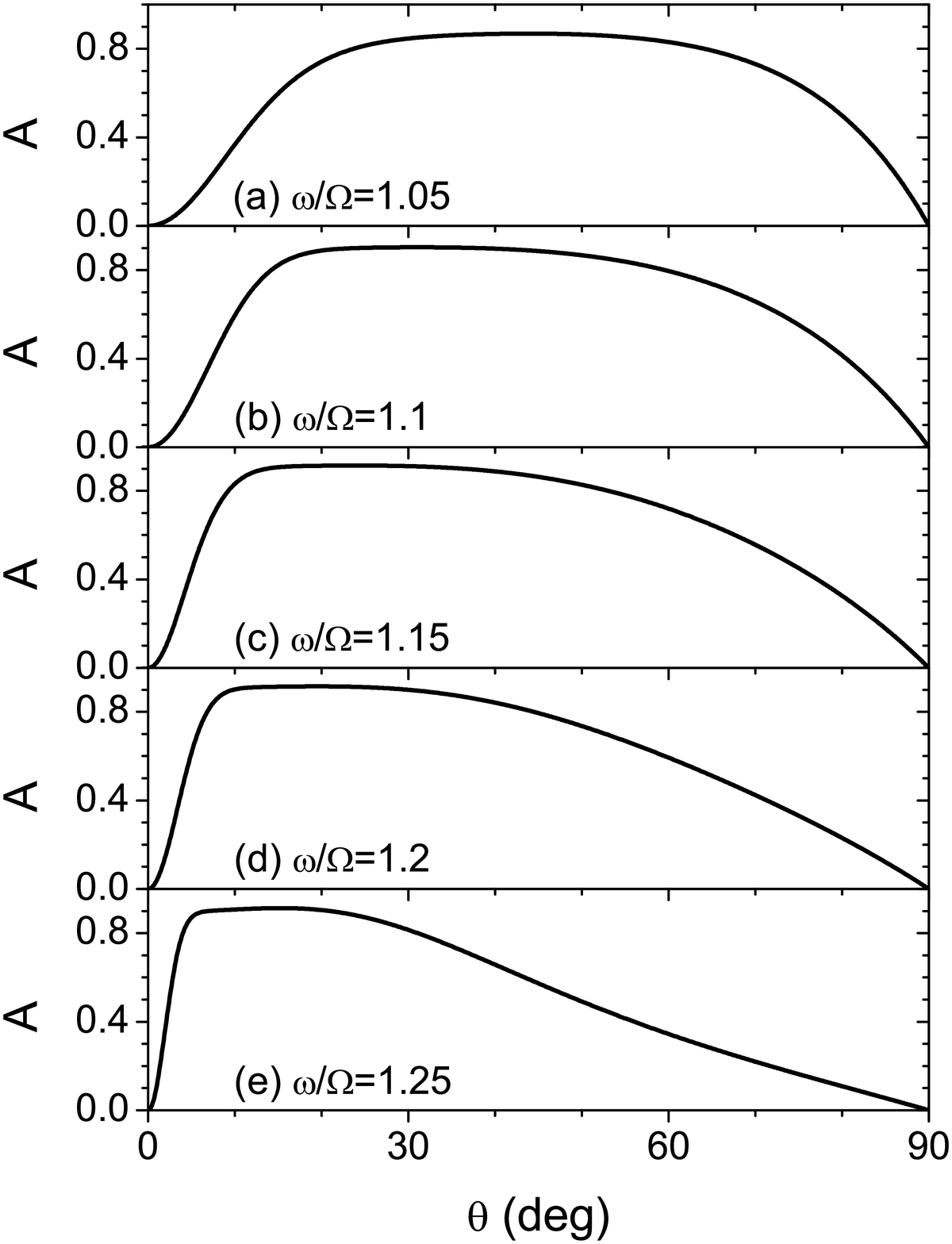}
\caption{Mode conversion coefficient $A$ plotted versus the incident angle $\theta$ for $\omega/\Omega=1.05$, 1.1, 1.15, 1.2 and
1.25, when $u=0.6$, $\eta=10^{-8}$, $L/\Lambda=50$ and $\Omega\Lambda/c=2.8$. }
\end{figure}

\begin{figure}
\centering
\includegraphics[width=3.2in]{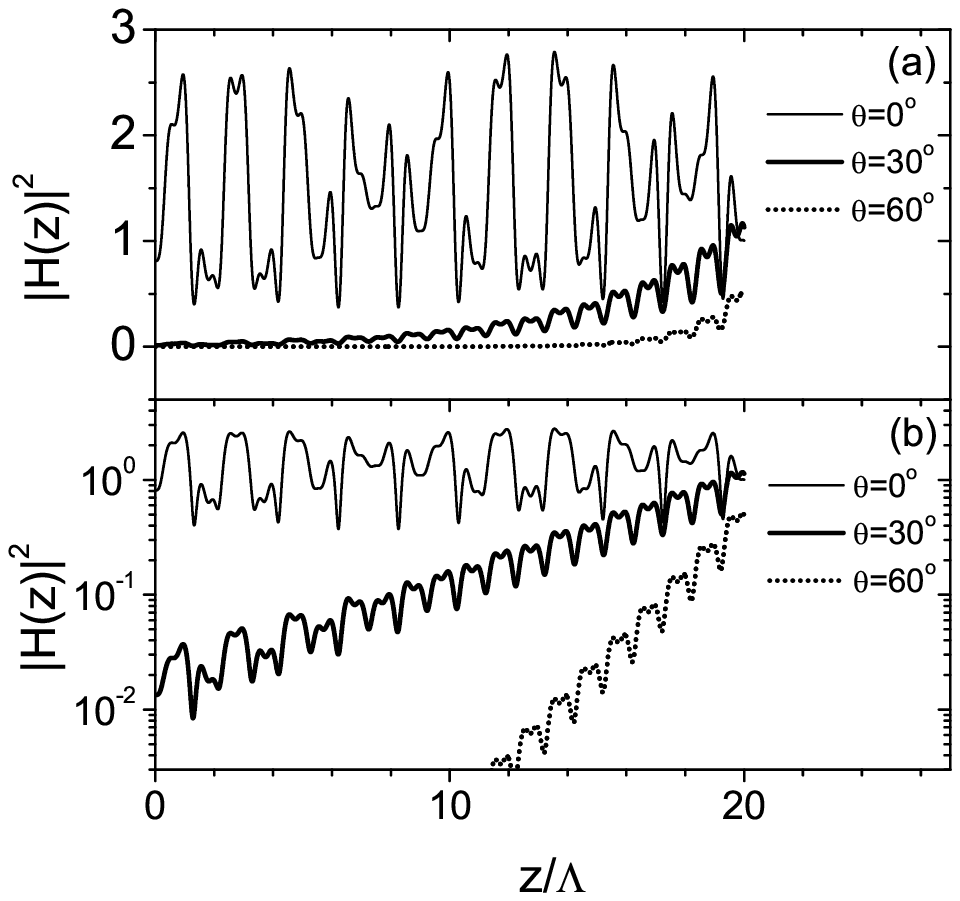}
\caption{Spatial distribution of the magnetic field intensity $\vert H(z)\vert^2$ for $\omega/\Omega=1.1$ and $\theta=0^\circ$, $30^\circ$ and $60^\circ$, when $u=0.6$, $\eta=10^{-3}$, $L/\Lambda=20$ and $\Omega\Lambda/c=2.8$, shown in (a) linear and (b) logarithmic plots.}
\end{figure}

In Fig.~4, we plot the mode conversion coefficient versus incident angle for $\omega/\Omega=1.05$, 1.1, 1.15, 1.2 and
1.25, when $u=0.6$, $\eta=10^{-8}$, $L/\Lambda=50$ and $\Omega\Lambda/c=2.8$. Again, we find that a substantial amount of mode conversion occurs for a wide range of the incident angle. For instance, when $\omega/\Omega=1.1$, $A$ is larger than 0.5 for $8.7^\circ<\theta<77^\circ$ and larger than 0.7 for $11.8^\circ<\theta<67.5^\circ$. This result is in sharp contrast to that for the case with a single resonance point, where the mode conversion occurs only for a rather narrow range of incident angle and the maximum value of the mode conversion coefficient is just 0.5 \cite{18,19}. Combining the results of Figs.~3 and 4, we can conclude that a broadband and wide-angle absorption enhancement is achieved due to the interplay between the mode conversion and the formation of a transmission band in PPCs.

Finally, in Fig.~5, we show the spatial distributions of the magnetic field intensity $\vert H(z)\vert^2$ inside the inhomogeneous plasma for $\omega/\Omega=1.1$ and $\theta=0^\circ$, $30^\circ$ and $60^\circ$, when $u=0.6$, $\eta=10^{-3}$, $L/\Lambda=20$ and $\Omega\Lambda/c=2.8$ in linear and logarithmic plots. We note that the wave is incident from the side where $z/\Lambda>20$. When $\theta$ is zero, no mode conversion occurs and the field distribution shows the characteristic of an extended state in the transmission band. The field intensity is not attenuated as the wave propagates toward $z=0$. When $\theta=30^\circ$, the mode conversion coefficient is very large ($A\approx 0.9$). Even though the overall field intensity decays exponentially with some periodical variation superimposed on it, it remains to be non-negligible close to $z=0$. This implies that the mode conversion and the associated resonant absorption occur at all 40 resonance points where $\epsilon=0$, thereby making the value of $A$ very large. When $\theta=60^\circ$, the mode conversion coefficient is smaller ($A\approx 0.796$). We observe that the field intensity decays faster than when $\theta=30^\circ$, which makes a less number of resonance points participate in the mode conversion process, thereby making $A$ smaller.

\section{Conclusion}
\label{sec5}

In this paper, we have performed a theoretical study of the mode conversion and the resonant absorption of $p$ waves in cold, unmagnetized and stratified plasmas with sinusoidal density variations. We have calculated the mode conversion coefficient and the spatial distribution of the magnetic field intensity as a function of the frequency and the incident angle using the invariant imbedding method. We have found that in contrast to the case where there is only one resonance point, the mode conversion coefficient in the periodic case can be significantly enhanced over a wide range of frequency and incident angle, which is due to the interplay between the mode conversion and the formation of a transmission band in PPCs. We expect that analogous phenomena will occur in more complicated plasmas such as magnetized plasmas with electrons and ions. The enhanced absorption of electromagnetic wave energy in such cases may be quite useful for an efficient heating of the plasma.

\acknowledgments
This work has been supported by the National Research Foundation of Korea Grant (NRF-2015R1A2A2A01003494) funded by the Korean Government.

\end{document}